\pdfoutput=1

\documentclass[11pt]{article}

\usepackage[dvipsnames]{xcolor}
\usepackage{ACL2023}

\usepackage{times}
\usepackage{latexsym}
\usepackage{comment}

\usepackage{array,float}
\usepackage{graphicx}
\usepackage{amssymb}
\usepackage{booktabs}
\usepackage{multirow}
\usepackage{enumitem}
\usepackage{xspace}
\usepackage[most]{tcolorbox}
\usepackage[capitalise]{cleveref}

\usepackage[T1]{fontenc}

\usepackage[utf8]{inputenc}

\usepackage{microtype}

\newcommand{\evalname}{\textsc{Sphere}\xspace}

\newcommand*\what{\color{BrickRed}}
\newcommand*\how{\color{orange}}
\newcommand*\who{\color{ForestGreen}}
\newcommand*\when{\color{cyan}}
\newcommand*\metahow{\color{violet}}

\title{\raisebox{-0.25\height}{\includegraphics[width=1cm]{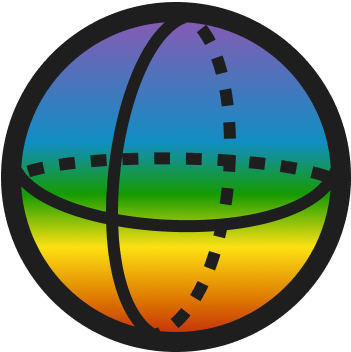}}\hspace{0.5em}\evalname: An Evaluation Card for Human-AI Systems}

\author{
\bf
Qianou Ma$^{1}$\thanks{ \ \ Co-first authors.} , 
Dora Zhao$^{2*}$, 
Xinran Zhao$^{1}$, 
Chenglei Si$^{2}$, 
Chenyang Yang$^{1}$, 
\\
\bf 
Ryan Louie$^{2}$, 
Ehud Reiter$^{3}$, 
Diyi Yang$^{2}$\thanks{ \ \ Co-last authors.} , 
Tongshuang Wu$^{1\dagger}$ \\ \\
$^1$Carnegie Mellon University, Pittsburgh, USA, \\
$^2$Stanford University, Stanford, USA, \\
$^3$University of Aberdeen, Aberdeen, UK
}

\begin{document}
\maketitle
\begin{abstract}
In the era of Large Language Models (LLMs), establishing effective evaluation methods and standards for diverse human-AI interaction systems is increasingly challenging. To encourage more transparent documentation and facilitate discussion on human-AI system evaluation design options, we present an evaluation card \evalname, which encompasses five key dimensions: 1) What is being evaluated?;~2) How is the evaluation conducted?;~3) Who is participating in the evaluation?;~4) When is evaluation conducted?;~5) How is evaluation validated? We conduct a review of 39 human-AI systems using \evalname, outlining current evaluation practices and areas for improvement. We provide three recommendations for improving the validity and rigor of evaluation practices.
\end{abstract}

{\centering
\begin{table*}[!htbp]
\footnotesize
    \caption{\small \evalname covers five dimensions of human-AI system evaluation, with 8 categories and 18 aspects. We provide examples and visualize the distribution of \evalname aspects from papers published at HCI (left bar) and NLP venues reviewed in our literature survey (details in \cref{sec:takeaways}).}
    \vspace{-5pt}
    \label{tab:taxonomy}
    \resizebox{\linewidth}{!}{

\begin{tabular}{ p{0.08\textwidth} p{0.086\textwidth} p{0.365\textwidth} p{0.02\textwidth} p{0.36 \textwidth}} 
\toprule
 \textbf{Category} & \textbf{Aspect} & \textbf{HCI Examples} & & \textbf{NLP Examples} \\
 
\midrule
\multicolumn{4}{l}{\textbf{{\what{What}} is being evaluated?}} \\
\midrule
     \multirow{2}{4em}{Component}    
            & Model     & Accuracy of LLM gen. \cite{lee_paperweaver:_2024} 
                        & \includegraphics[width=\linewidth,height=0.3cm,trim={2cm 3.7cm 40.5cm 0.5cm},clip]{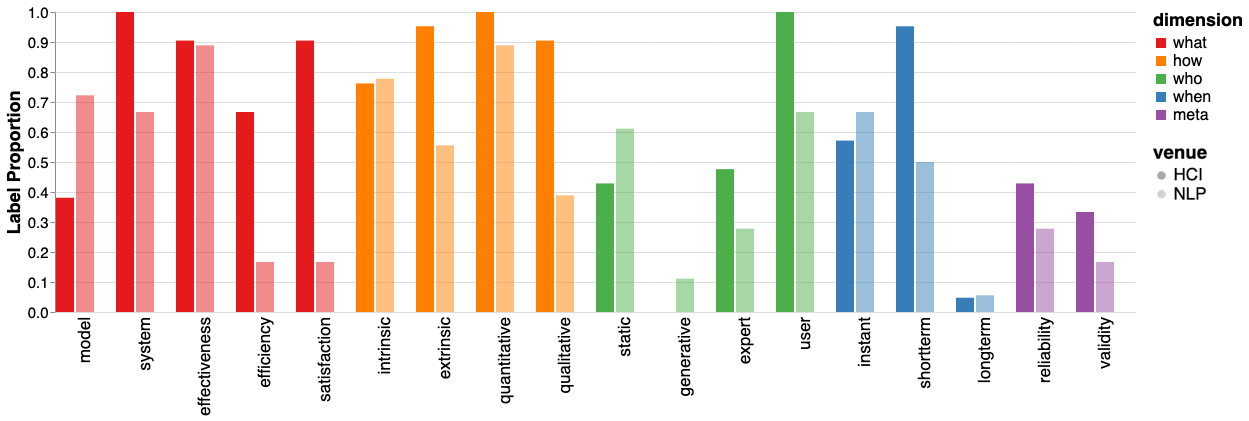}
                        & BERTScore~\cite{gloria-silva_plan-grounded_2024} 
                        \\ 
            & System    & Knowledge quiz \cite{shaikh_rehearsal:_2024} 
                        & \includegraphics[width=\linewidth,height=0.3cm,trim={4cm 3.7cm 38.5cm 0.5cm},clip]{tab_fig/labels.png}
                        & Headline quality \cite{ding_harnessing_2023} 
                        \\
     \multirow[t]{3}{4em}{Design Goal} 
        & Effectiveness & System risk \& content~\cite{rajashekar_human-algorithmic_2024} 
                        & \includegraphics[width=\linewidth,height=0.3cm,trim={6.2cm 3.7cm 36.5cm 0.5cm},clip]{tab_fig/labels.png}
                        & Label quality \& stability \cite{wei_collabkg:_2024}
                        \\

        & Efficiency    & Perceived workload \cite{lee_paperweaver:_2024} 
                        & \includegraphics[width=\linewidth,height=0.3cm,trim={8.5cm 3.7cm 34.5cm 0.5cm},clip]{tab_fig/labels.png}
                        & Time on task \cite{ding_harnessing_2023} 
                        \\ 
                        
        & Satisfaction  & Likert-scale rating of fun \cite{wang_virtuwander:_2024} 
                        & \includegraphics[width=\linewidth,height=0.3cm,trim={10.5cm 3.7cm 32.3cm 0.5cm},clip]{tab_fig/labels.png}
                        & Likert-scale rating of trust \cite{ding_harnessing_2023} 
                        \\ 

\midrule
\multicolumn{4}{l}{\textbf{{\how{How}} is an evaluation conducted?}} \\
\midrule
     Scope & Intrinsic  & System Usability Scale \cite{liu_selenite:_2024}
                        & \includegraphics[width=\linewidth,height=0.3cm,trim={12.6cm 3.7cm 30.2cm 0.5cm},clip]{tab_fig/labels.png}
                        & Retrieval hit rate \cite{inan_generating_2024} 
                        \\ 
           & Extrinsic  & Engagement \& enjoyment \cite{fan_contextcam:_2024} 
                        & \includegraphics[width=\linewidth,height=0.3cm,trim={14.7cm 3.7cm 28.1cm 0.5cm},clip]{tab_fig/labels.png}
                        & Identified concept diversity \cite{yang_beyond_2023} 
                        \\ 
     Method & Quantitative  & Interaction logs~\cite{wu_ai_2022} 
                            & \includegraphics[width=\linewidth,height=0.3cm,trim={16.8cm 3.7cm 25.8cm 0.5cm},clip]{tab_fig/labels.png}
                            & Micro F1 \cite{wei_collabkg:_2024} 
                            \\ 
            & Qualitative   & Interview \& grounded coding \cite{liu_what_2023} 
                            & \includegraphics[width=\linewidth,height=0.3cm,trim={18.9cm 3.7cm 23.8cm 0.5cm},clip]{tab_fig/labels.png}
                            & Case study \cite{cai_low-code_2024} 
                            \\ 
                            
\midrule
\multicolumn{4}{l}{\textbf{{\who{Who}} is participating in the evaluation?}} \\
\midrule
     Human     & Expert     & Prolific experts \cite{zavolokina_think_2024} 
                            & \includegraphics[width=\linewidth,height=0.3cm,trim={25.2cm 3.7cm 17.4cm 0.5cm},clip]{tab_fig/labels.png}
                            & ASL expert \cite{inan_generating_2024} 
                            \\  
               & User       & Students \& physicians \cite{rajashekar_human-algorithmic_2024} 
                            & \includegraphics[width=\linewidth,height=0.3cm,trim={27.4cm 3.7cm 15.4cm 0.5cm},clip]{tab_fig/labels.png}
                            & Crowdworkers \cite{chakrabarty_help_2022} 
                            \\
     \multirow{2}{4em}{Automated}
               & Static     & Perplexity \& LIWC scores \cite{calle_towards_2024} 
                            & \includegraphics[width=\linewidth,height=0.3cm,trim={21cm 3.7cm 21.8cm 0.5cm},clip]{tab_fig/labels.png}
                            & Precision \& recall \cite{yang_beyond_2023} 
                            \\ 
               & Generative & N/A 
                            & \includegraphics[width=\linewidth,height=0.3cm,trim={23.2cm 3.7cm 19.6cm 0.5cm},clip]{tab_fig/labels.png}
                            & Consistency by LLaMa2 \cite{zhao_narrativeplay:_2024} 
                            \\
                            
\midrule
\multicolumn{4}{l}{\textbf{{\when{When}} is evaluation conducted (duration)?}} \\
\midrule
    \multirow[t]{3}{4em}{Time Scale} 
        & Immediate   & \# of clicks \cite{lawley_val:_2024} 
                    & \includegraphics[width=\linewidth,height=0.3cm,trim={29.4cm 3.7cm 13.2cm 0.5cm},clip]{tab_fig/labels.png}
                    & Benchmark \cite{raheja_coedit:_2023} 
                    \\ 
        & Short-term    & 1-hour usability study \cite{liu_selenite:_2024} 
                        & \includegraphics[width=\linewidth,height=0.3cm,trim={31.6cm 3.7cm 11cm 0.5cm},clip]{tab_fig/labels.png}
                        & 10 minutes per poem \cite{chakrabarty_help_2022} 
                        \\
        & Long-term     & 3-days session \cite{fan_contextcam:_2024} 
                        & \includegraphics[width=\linewidth,height=0.3cm,trim={33.8cm 3.7cm 9.1cm 0.5cm},clip]{tab_fig/labels.png}
                        & 6-months deployment \cite{inan_generating_2024} 
                        \\

\midrule
\multicolumn{4}{l}{\textbf{{\metahow(Meta) How} is evaluation validated?}} \\
\midrule
    Validation 
         & Reliability  & Krippendroff's $\alpha$ as IRR \cite{lee_paperweaver:_2024} 
                        & \includegraphics[width=\linewidth,height=0.3cm,trim={35.8cm 3.7cm 6.9cm 0.5cm},clip]{tab_fig/labels.png}
                        & Fleiss' $\kappa$ for annotation \cite{zhao_narrativeplay:_2024} 
                        \\ 
        & Validity      & Counterbalancing \cite{wu_ai_2022} 
                        & \includegraphics[width=\linewidth,height=0.3cm,trim={38cm 3.7cm 4.7cm 0.5cm},clip]{tab_fig/labels.png}
                        & Randomized control \cite{ding_harnessing_2023} 
                        \\ 
\bottomrule  
\end{tabular}
}
\vspace{-10pt}
\end{table*}
}
\section{Introduction}
The proliferation of LLMs has changed the way humans interact with AI systems. Compared to previously existing AI models, LLMs can better comprehend and generate human-like text, enabling users to engage with AI through natural language in a more conversational manner \cite{brown2020gpt3}. By leveraging these capabilities, system designers can create human-AI systems\textsuperscript{\ref{footnote:haisys-def}} that span a range of domains and roles. Despite the rapid advances in designing new human-AI systems, it is still unclear how to best evaluate them. Standard evaluation practices in NLP are better suited for understanding model performance using automated metrics and static benchmarks. While there have been efforts to integrate more human-centered evaluation, there are also concerns about the validity and reproducibility of how these evaluations are conducted~\cite{belz2023missing,howcroft-etal-2020-twenty,gehrmann2023repairing}.

Given the wide diversity of human-AI systems, what factors should researchers consider when designing evaluations? How do we ensure that these evaluations are transparent and replicable? To address these questions, {we need systematic methods for documenting how these evaluations are conducted. As a step in this direction,} we propose the \emph{\evalname} evaluation card, which provides a comprehensive template for designing and documenting evaluation protocols used to assess human-AI systems.\footnote{Following the definitions \label{footnote:haisys-def} from the literature \cite{lee2022evaluating, amershi2019GuidelinesHAIInteraction}, a human-AI system harnesses AI capabilities that are exposed to end users through an interface. These systems consist of an AI model, a user interface, and logic that converts user entry into input for the model. We instantiate the AI as LLM when we refer to human-AI systems and evaluation in this paper.} Although we focus on systems powered by LLMs in this work, the evaluation dimensions we discuss are agnostic to the type of model and can be applied to AI systems more broadly.

\begin{tcolorbox}[
    squeezed title*={\textbf{\evalname} Evaluation Card for Human-AI Systems},
    fonttitle=\small,
    colframe=gray,
    colback=white,
    subtitle style={boxrule=0pt, colback=black!5!white, colupper=black!75!gray},
    fontupper=\footnotesize 
]

\tcbsubtitle{ (\textbf{\underline{S}}ubject) {\what\textbf{What}} is being evaluated? }
\begin{itemize}[leftmargin=0pt,nosep,itemsep=2pt]
    \item \textbf{Component}: Model, System
    \item \textbf{Design Goal}: Effectiveness, Efficiency, Satisfaction
\end{itemize}

\tcbsubtitle{(\textbf{\underline{P}}rocess) {\how\textbf{How}} is the evaluation being conducted? }
\begin{itemize}[leftmargin=0pt,nosep,itemsep=2pt]
    \item \textbf{Scope}: Intrinsic, Extrinsic
    \item \textbf{Method}: Quantitative, Qualitative
\end{itemize}

\tcbsubtitle{(\textbf{\underline{H}}andler) {\who\textbf{Who}} is participating in the evaluation? }
\begin{itemize}[leftmargin=0pt,nosep,itemsep=2pt]
    \item \textbf{Automated}: Static, Generative
    \item \textbf{Human}: Expert, User
\end{itemize}

\tcbsubtitle{(\textbf{\underline{E}}lasped) {\when\textbf{When}} is evaluation conducted (duration)? }
\begin{itemize}[leftmargin=0pt,nosep,itemsep=2pt]
    \item \textbf{Time Scale}: Immediate, Short-term, Long-term
\end{itemize}

\tcbsubtitle{(\textbf{\underline{R}}obustness) {\metahow\textbf{How}} is evaluation validated?}
\begin{itemize}[leftmargin=0pt,nosep]
    \item \textbf{Validation}: Reliability, Validity
\end{itemize}
\end{tcolorbox}

The purpose of \evalname is two-fold. First, researchers can use the template when \emph{designing} evaluations (\cref{sec:framework}). We enumerate five high-level questions that scaffold the dimensions that researchers should think through to ensure that their evaluation aligns with the intended design goal. Within each dimension, we discuss popular practices from both NLP (e.g., benchmarking, LLM-as-a-judge) and from HCI (e.g., experimental user studies, semi-structured interviews) that can be adopted. Second, \evalname can be used to \emph{document} how evaluation is conducted. To address concerns around the replicability and reproducibility of results, researchers can use the cards to communicate their evaluation design. This practice helps improve transparency in the field and fosters a shared language for researchers to communicate even if the systems being evaluated are different
(see examples in \cref{appen:case_study}).\footnote{{Template for \evalname is released on \href{https://sphere-eval.github.io/}{sphere-eval.github.io}.}}

Using \evalname, we analyze 39 human-LLM systems from NLP and HCI venues (\cref{sec:takeaways}). From our analysis, we provide three key recommendations for improving evaluation practices:~1) establish evaluations in real-world contexts;~2) strengthen validity and interpretability of results via triangulating various evaluation methods; and~3) rigorously evaluate evaluation practices. These recommendations bridge the strengths of both communities, for example, HCI's focus on stakeholder relevance and NLP's advancements in automatic quantitative measures, ultimately shaping more robust and actionable evaluation.
Finally, we present two case studies (\cref{sec:casestudy}) showcasing \evalname for designing and reproducing evaluations.

\section{Background}
NLP is facing what some scholars have termed an ``evaluation crisis'' --- how to best evaluate the capabilities of generative models remains an open question~\cite{blodgett-etal-2024-human,xiao2024human}. Established NLP methods, such as static benchmarks, are found to be ill-suited to judge model performance on generative tasks~\cite{mcintosh2024inadequacies}, spurring efforts to make more dynamic and comprehensive benchmarks~\cite{liang2022holistic}. Others have advocated for more \emph{human}-centered evaluations of model capabilities~\cite{liao2023rethinking,blodgett-etal-2024-human,elangovan2024considers}. These concerns also coincide with alarms around the lack of experimental reproducibility and repeatability in human evaluations of NLP systems~\cite{belz2023missing}. Taken together, there is growing uncertainty about how evaluations should be conducted going forward and how to ensure the quality of evaluation results.

Going beyond model capabilities, evaluating human-AI systems introduces additional challenges. Researchers must consider not only the model performance but also the system's impact on users~\cite{weidinger2023sociotechnical}. Although there are many guidelines about how to design human-AI systems~\cite{amershi2019GuidelinesHAIInteraction,wright2020comparative}, there is less literature on how we should be evaluating them. Examples of work that tackle system evaluation include \citet{lee2022evaluating}'s framework for human-LLM interaction evaluation that captures the process and user preferences beyond static model outputs quality. Others have also proposed methods focused on assessing the safety of these systems~\cite{ibrahim2024beyond,weidinger2023sociotechnical}, or for domain-specific applications~\cite{lee2024design}. However, there is still a gap in articulating a comprehensive overview of how to design evaluations for human-AI systems.

To address this, we present the \evalname evaluation card: a framework covering five dimensions of human-AI system evaluation that helps researchers \emph{design} and \emph{document} evaluation. 
As a {design} tool, \evalname structures conversations around key evaluation areas. As {documentation}~\cite{gebru2021datasheets,mitchell2019model}, \evalname contributes to the transparency and reproducibility of these methods.

\section{The \evalname Evaluation Card}
\label{sec:framework}

In this section, we present the five dimensions and corresponding aspects of evaluation included in our \evalname evaluation card (\cref{tab:taxonomy}), and we highlight the challenges and considerations for each dimension.
Similar to existing documentation efforts \cite{mitchell2019model}, we note that these dimensions are not intended to be exhaustive. Researchers may want to report additional dimensions of evaluation depending on their system design. {See~\cref{fig:anglekindling} for an example \evalname card.}

\begin{figure}
    \centering
    \input{tab_fig/anglekindling_simplified}
    \vspace{-10pt}
    \caption{Example \evalname card for the system AngleKindling~\cite{petridis2023anglekindling}.} 
    \label{fig:anglekindling}
    \vspace{-5pt}
\end{figure}

{
\paragraph{Method} We develop our framework using an expert-based affinity diagramming approach \cite{hartson2012ux,lucero2015using,harboe2015real} across 9 authors.\footnote{Authors have a median of 4 years experience in human-AI interaction and system evaluation in NLP and HCI.} Each author enumerated important aspects and theories of human-AI system evaluation in their domains. Via synchronous conversations, we clustered these aspects into high-level themes, prioritizing the most salient ones. We then organized the themes using the Who, What, When, and How questions, taking inspiration from how these general dimensions have guided exploration and analysis across domains~\cite{apte20015w}. Finally, we refined the dimensions after applying them to two example systems (see \cref{appen:case_study}).
}

\subsection{{\what\textbf{What}} is being evaluated?}
\label{subsec:what}

The first question to answer when designing an evaluation is to determine \emph{what} is being evaluated. To answer this, we discuss two categories of aspects: which part of the system is the focus of the evaluation and what goal the evaluation is testing.

\paragraph{Components} Since human-AI systems consist of multiple components, we must identify what part of the system is being evaluated. A helpful delineation is between evaluating \textbf{model} behavior versus the \textbf{system} as-a-whole.

We break out the \textbf{model} as a separate component in evaluation since it represents a unique design challenge of human-AI systems, driven by uncertainty surrounding model capabilities and the complexity of outputs~\cite{yang2020re}. Uncertainty can be compounded by the fact that systems may include multiple models with different functions (e.g.,~\citet{wang_virtuwander:_2024}). 

Yet, models are but one part of a more complex artifact; designers introduce interfaces and interactions that integrate with the model components to form a human-AI \textbf{system}. Understanding how these design choices impact users' experiences requires evaluating the system holistically.

\paragraph{Design Goals} {Evaluations should be formulated to help prove a design goal of the model or system.} To taxonomize possible design goals, we use three categories from the ISO standard definition of usability~\cite{bevan2016new}:
\begin{itemize}[topsep=0pt,itemsep=-1ex,partopsep=1ex,parsep=1ex,leftmargin=5pt]
    \item \textbf{Effectiveness}: accuracy, completeness, and lack of negative consequences with which users achieved specified goals. This category maps closely to the quality criteria used in NLG evaluation~\cite{howcroft-etal-2020-twenty, reiter2024nlg}.
    \item \textbf{Efficiency}: resources (e.g., time, cognitive effort) required to achieve the {model's or system's goals}.
    \item \textbf{Satisfaction}: positive attitudes, emotions, and/or comfort resulting from the use.
\end{itemize}

\paragraph{Considerations: selecting design goals and scenarios}
We provide two considerations for deciding what to evaluate. First, human-AI systems do not have to aim for all three design goals. In fact, researchers should consider how designing a system for one goal could harm another. For example, systems relying on entirely automated decision-making may be more efficient but can be considered less trustworthy~\cite{hong2020human}.

Second, researchers should factor in how goals differ across scenarios. System performance will vary depending on whether we are evaluating the ``average'' versus worst case. Particularly in high-stakes domains, we must consider this long-tail of system behavior as they may pose immense harm~\cite{Bickmore:jmir18}. One popular technique is red-teaming systems, although finding adversarial scenarios may be expensive and hard to identify pre-deployment~\cite{ganguli2022red,mei-etal-2023-assert}. Once deployed, the system should be monitored for failures, requiring corrective action or even removal of the system in extreme cases~\cite{syed2015black,wolf2017we}.

\subsection{{\how\textbf{How}} is the evaluation conducted?}
\label{subsec:how}
Next, we must consider how the evaluation is conducted, including the scope of the evaluation and methods used. Note that human-AI systems may require multiple evaluations employing different scopes and methods. For example, researchers may validate the model or interface design before evaluating the system with users. Each evaluation would be conducted differently. 

\paragraph{Scope}
Evaluating human-AI systems requires assessing both their internal capabilities (\textbf{intrinsic evaluation}) and performance in real-world scenarios (\textbf{extrinsic evaluation})~\cite{jones1995evaluating}. Prior work has pointed out that NLP systems tend to disproportionately favor intrinsic evaluation, and have pushed for more extrinsic evaluation~\cite{gkatzia2015snapshot,gehrmann2023repairing}. While intrinsic evaluations are still important, researchers should be mindful of how well these internal metrics correlate to real-world utility~\cite{reiter2009investigation,belz2008intrinsic}.

\paragraph{Method}
Both \textbf{quantitative} and \textbf{qualitative} methods can be used to conduct intrinsic and extrinsic evaluations but yield different types of insights and can be complementary to each other. In mixed-method analyses, qualitative methods add more nuance to quantitative results. For example, \citet{petridis2023anglekindling}'s quantitative results establish that their proposed system outperforms an existing one, while qualitative responses in a semi-structured interview unearth reasons why their system failed and highlight possible improvements. 

\paragraph{Considerations: selecting and implementing evaluation methods }
Different methods have their own considerations to ensure rigorous execution. Quantitative analysis provides generalizable insights and facilitates comparison across groups. Therefore, when using quantitative methods, researchers should consider selecting metrics and datasets that are representative of the system's task (e.g., when benchmarking) or sampling participants to ensure experiments have sufficient statistical power~\cite{charness2012experimental}. On the other hand, qualitative analysis aims to provide a deeper understanding or ``thick description''~\cite{geertz2008thick} of behavior. Generalizability may \emph{not} be a priority~\cite{donmoyer2000generalizability,soden2024evaluating}. Researchers may want to consider extant theory, sampling strategies, and coding techniques (e.g., open coding, selective coding)~\cite{cairns2008research,cole2022more}.

\subsection{{\who\textbf{Who}}  is participating in the evaluation?}
\label{subsec:who}
When designing evaluation methods, researchers must consider who (or what) is participating in this evaluation, including human-centered evaluations and automated methods spanning from standard benchmarking techniques to LLM judges.

\subsubsection{Human Evaluators}
When selecting human evaluators, we must account for how participants' identities or backgrounds may influence how they interact with human-AI systems. HCI work has provided frameworks for thinking about how demographic background~\cite{bardzell2011towards, ogbonnaya2020critical, schlesinger2017intersectional} influences how individuals may interact with technology. In addition to these factors, in our \evalname evaluation cards, we highlight two {aspects}: whether participants are the intended design targets and their level of expertise. 

\paragraph{Intended Users}
When selecting human evaluators, we must first decide, whether the system is evaluated by the user (i.e., the intended design target) or another stakeholder. While responses from intended users will more closely mirror how people will interact with the deployed system, {there are other stakeholders who may be affected by the system and whose input should be considered. For example, \citet{ma2024hypocompass} had teachers evaluate an AI-infused tutoring system, for which students are the intended users; other potential evaluators could have been parents or administrators. {Working across stakeholder groups can lead to tensions when communities may have differing priorities or standards for evaluation~\cite{reiter2009investigation}.}

\paragraph{Experts} {Another point to consider is the expertise of the evaluators. Expertise can refer to the domain expertise, such as having trained physicians evaluate a clinical decision support system~\cite{rajashekar_human-algorithmic_2024} and having teachers evaluating a tutoring system \cite{ma2024hypocompass}. Here, expertise is not to be confused with experts from crowdworking platforms, which can also refer to workers who have passed qualification studies~\cite{chakrabarty_help_2022}. Domain expert evaluations can produce more reliable judgments than non-experts~\cite{yesilada2009much} and bring field-specific insights, but their perceptions may not align with those of the intended system users. Working with expert evaluators can also be more expensive and time-intensive.}

\subsubsection{Automated Evaluators}
There are many ways to automatically evaluate systems. We point out two main categories: evaluations where outputs are compared to an established reference and evaluations using generative models to make judgments~\cite{zheng2024judging}.

\paragraph{Static Evaluators} We refer to methods that compare model or system behaviors to some existing ground-truth behavior as static evaluation. For example, benchmarking with perplexity metrics or rule-based evaluations~\cite{blagec2022global,van2020evaluation} fall in this category. Prior work has raised concerns about the usefulness and validity of existing benchmarks for assessing generative tasks~\cite{mcintosh2024inadequacies}. Benchmarking may not be necessary when systems involve less model implementation, such as when researchers prompt an off-the-shelf model.

\paragraph{Generative Evaluators}
Researchers can also use LLMs to judge outputs (typically originating from another LLM).\footnote{We would not consider an embedding-based evaluation method such as BERTScore~\cite{zhangbertscore} as a generative evaluation. The method compares BERT embeddings between the output and a reference rather than using the model for generation.} For example, \citet{zhao_narrativeplay:_2024} use LLaMA-2-70B to rate the system's responses along the dimensions of consistency, relevance, empathy, and commonsense. 
Generative evaluators allow researchers to run more ablations under controlled conditions and iterate on system design quickly~\cite{zheng2024judging}. Generative evaluators operate under the premise that the models' choices are similar to that of humans; however, there remain concerns about biases inherent in dataset construction methods~\cite{wang2023pandalm} or preferences for outputs generated from certain models~\cite{li2023collaborative,yin2023large}.

\paragraph{Considerations: specifying relevant evaluators}
When deciding who to include as evaluators, researchers should aim for specificity. Some human-AI systems are presented as ``general-purpose'' or without a defined intended user group. Works like Chatbot Arena~\cite{chiang2024chatbotarena} open up system evaluation to an indiscriminate potential end-user. Nonetheless, there is still value in recruiting specific user groups, such as participants who may be historically excluded from the development of such technologies~\cite{ogbonnaya2020critical} or users who might interact with the system in more high-risk domains~\cite{rauh2024gaps}. 

Similar concerns apply when using automated methods. Researchers should reflect on whose perspectives are excluded from the evaluation. Generative models may not adequately simulate diverse personas or capture the nuances of different identity groups~\cite{wang2024large,cheng2023compost}.
Benchmarks are also not immune; they are shaped by the design biases and positionality of the dataset creator~\cite{santy2023nlpositionality}.

\subsection{{\when\textbf{When}} is evaluation conducted (duration)?}
\label{subsec:when}
While many standard benchmark evaluations for AI systems can be run in seconds, evaluating human-AI systems requires us to consider time factors. Drawing from \citet{newell1985prospects}'s time scales of human action, we discuss evaluations that can occur at three different time scales:
\begin{itemize}[topsep=0pt,itemsep=-1ex,partopsep=1ex,parsep=1ex,leftmargin=5pt]
    \item \textbf{Immediate}: When evaluation occurs at the time-scale of milliseconds and seconds, rational thought processes are not yet at play~\cite{newell1985prospects}. For instance, telemetry or log data can be used to analyze real-time interactions~\cite{liu_how_2024, kim_evallm:_2024}. Similarly, automated benchmarking approaches measure performance at a fixed point in time. 
    \item \textbf{Short-term}: Evaluating over the course of minutes or hours sheds light on human behaviors and thoughts in a bounded context. Short-term evaluation is crucial for measuring the benefits of interacting with a human-AI system. However, they may be biased by known psychological phenomena, such as the novelty effect~\cite{elston2021novelty}, and fail to capture longer-term impacts of usage. 
    \item \textbf{Long-term}: Studies operating on longer time scales (days, months, years, and more) capture behavioral changes and effects from social interaction that may not appear in isolated laboratory experiments. Over time, users also may form different mental models of the systems, changing their interaction patterns. For example, \citet{bansal2019beyond} found that users' trust in AI systems evolved with prolonged exposure and interaction.
\end{itemize}

\paragraph{Considerations: balancing desired outcomes with practicalities of evaluation duration}

{
When deciding the duration of evaluation, researchers must weigh trade-offs between desired outcomes and practical factors. Immediate evaluations, such as automated methods, are cheap and efficient, but they may fail to capture the consequences of interacting with the system. Alternatively, longitudinal studies are crucial to understanding sustained impacts and broader implications of AI adoption in real-world workflows (e.g., privacy expectations~\cite{khowaja2024chatgpt}, workforce impact~\cite{microsoftfuture}). With new technologies, the novelty effect can bias short-term evaluation~\cite{long2024not}. Long-term evaluations, however, are time-consuming and financially expensive~\cite{caruana2015longitudinal}.  Researchers also need to manage high attrition rates and possibly intervene if drop-out follows systematic patterns~\cite{hogan2004handling}.}

\subsection{{\metahow\textbf{How}} is evaluation {validated}?}
\label{subsec:meta} 
Finally, researchers must ensure that their evaluations are sound and replicable. Drawing on concepts from the social sciences~\cite{bandalos2018measurement,drost2011validity}, we present two qualities --- reliability and validity --- that should be assessed when evaluating the evaluation design. We refer to this step as ``meta-evaluation.''

\paragraph{Reliability}
Researchers must consider reliability, or whether the evaluation produces consistent results. Three important dimensions of reliability are as follows: stability over time (``Do results remain the same across time points?''); equivalence (``Are results consistent across different versions of the evaluation?''); and internal consistency (``Do the components of my evaluation method measure the same concept?'')~\cite{drost2011validity}. For example, \citet{taeb_axnav:_2024} included three Likert-scale questions to evaluate system usefulness; to check internal consistency, we want to validate that the questions all measure the same concept. 

\paragraph{Validity} Measures can be reliable but still not be valid. Researchers must also consider whether their evaluation techniques meaningfully capture the intended construct \cite{adcock2001measurement-validity}. Assessing validity is particularly important when evaluating properties that are socially constructed (e.g., system's impact on creativity~\cite{he2023wordart,fan_contextcam:_2024}) compared to more objective measures (e.g., task completion time~\cite{liu_what_2023}). 

\paragraph{Considerations: {carefully executing} valid, reliable, and replicable evaluations}
For reliability, a popular measure of internal consistency is Cronbach's $\alpha$, which intuitively captures correlations between test items~\cite{tavakol2011making}. When humans or models are used to rate system outputs, researchers must account for the reliability of their judgments. For example, LLM judges can produce different responses even when given the same prompt~\cite{stureborg2024large,shi2024judging}. We can measure annotation consistency or inter-rater reliability (IRR) across different raters~\cite{artstein2008inter, gwet2001handbook}. Common metrics for IRR include Fleiss's $\kappa$ and Cohen's $\kappa$ for categorical labels, correlation metrics for continuous scores, and rank-based correlations, including Spearman's $\rho$ or Kendall's $\tau$ for rank labels. Methods such as test-retest and parallel measures of the same concept are also useful for assessing other aspects of reliability~\cite{drost2011validity}.

Depending on the evaluation, there are different considerations for validity. For behavioral experiments, researchers must mitigate systematic biases (e.g., confirmation bias~\cite{confirmation-bias}, anchoring bias~\cite{BLOCK1991188}) that can impact experimental results. For automated methods, particularly generative evaluators, model biases or errors may differ from those of humans~\cite{bavaresco2024llms}. Thus, researchers need to check the consistency between the automated results and human perception~\cite{gehrmann2023repairing, shankar2024validates}. {Overall, evaluation experiments must be carefully executed in order to be meaningful and provide real-world utility~\cite{reiter2024nlg}.}

\section{Applying the \evalname Evaluation Card}
\label{sec:takeaways}

\evalname provides a structured framework for analyzing and improving how human-AI systems are evaluated. In this section, we apply \evalname to works published in NLP and HCI venues. Through this analysis, we uncover trends, gaps, and best practices that might otherwise be overlooked, such as underrepresented evaluation aspects or core evaluation designs. We distill actionable lessons to guide researchers in designing more robust evaluation paradigms that align with the real-world contexts in which these systems operate.

\subsection{Method}
To identify human-LLM systems, we searched for papers published in human-computer interaction (CHI) or natural language processing (*CL) venues between Jan. 2022 and Sept. 2024. We kept papers that mentioned \texttt{{[“human*”} OR “user*”] AND ``system*'' AND [``large language model*'' OR ``LLM*'']} in the abstract or title and then manually inspected to ensure that a human-LLM system was introduced.\textsuperscript{\ref{footnote:haisys-def}} We reviewed 39 papers --- with 21 papers from HCI venues and 18 from NLP. 
Six of the authors analyzed the papers. We first independently coded two of the 39 papers (Krippendorf's $\alpha = 0.69$) before discussing disagreements synchronously. The remaining papers were then divided among the authors. We visualize the distribution of codes across papers from HCI and NLP venues in Table~\ref{tab:taxonomy}. See \cref{appen:method} for details.

\subsection{Recommendations} 

Building on insights learned in our analysis with \evalname, we propose three recommendations to improve the quality of human-AI system evaluations.

\subsubsection{Evaluate to reflect real-world use} 
A core challenge of evaluating human-AI systems is bridging the gap between model benchmark performance and real-world usage. Extrinsic evaluation is critical for understanding how systems will perform in the real world. Our analysis found that only 10 of the 18 NLP papers included extrinsic evaluation compared to 20 of the 21 HCI papers. 

\paragraph{Test systems in the real world} The most common method across the 30 papers is running within- or between-subjects experiments to quantitatively compare systems for a pre-determined task. One issue with this paradigm is that it is limited to controlled laboratory settings, which may lack ecological validity. Results may not always translate to real-world scenarios.
Only two of the reviewed papers deployed the system to the real world and studied user behavior in situ. \citet{fan_contextcam:_2024} had participants use the ContextCam system in their day-to-day lives over three days, and \citet{inan_generating_2024} had public users use their multimodal dialogue system for over six months. {Going beyond general public deployment, researchers can consider evaluating their systems in users' real-world workflows, providing deeper insight into the system's utility, particularly for professional contexts (e.g.,~\citet{knoll2022user}).} Real-world deployment gives insight into how actual users perceive the system in a realistic setting and across contexts --- the ultimate test for human-AI systems.

\paragraph{Recruit evaluators from relevant stakeholder groups} Selecting evaluators is crucial to ensure population validity. Researchers should recruit relevant stakeholders when running human evaluations. Only three of the 13 NLP papers reviewed with user evaluations recruited domain experts. Others did not provide details on user background or relied on crowdworkers and students who may not represent the target user population. 
Using crowdworkers or convenience sampling can be more time and cost-efficient. However, downsides include higher variance in responses and a potential lack of relevant expertise for domain-specific evaluation~\cite{karpinska2021perils}. 

Finding the right evaluators is challenging. One way to recruit more representative users is through collaborating with relevant organizations. For example, when evaluating their mental health counselor training system, \citet{hsu2023helping} worked with 7 Cups of Tea, an existing platform for online mental health support, to recruit participants. Examining evaluation practices has implications upstream regarding how systems should be designed. Working more closely with users during the design process, such as conducting formative studies or adopting co-design practices, better motivates the system and forms relationships for finding evaluators~\cite{burkett2012introduction}. There has also been growing interest in using LLMs to simulate human raters or even using human-LLM collaboration for annotation \cite{li2023collaborative,Gao2024nlg}. As addressed in \cref{subsec:who} and \cref{subsec:meta}, if adopting these methods, we recommend researchers first evaluate whether the generated responses align with users.

\subsubsection{Cross-verify results across different evaluation methods}
With the vast space of applications and interactions that human-AI systems afford, {we argue that} it is important to adopt different methods to increase the evaluation robustness and cross-verify the findings. However, less than half of the papers in our corpus utilized all intrinsic, extrinsic, quantitative, and qualitative methods to cross-validate their evaluation results. When using the \evalname card to design evaluations, we advocate researchers pay attention to whether a multifaceted evaluation is adopted to help mitigate biases inherent in any single evaluation method (\cref{subsec:how}).

\paragraph{Triangulate results} To improve confidence in evaluation results, methodological triangulation --- or using more than one method for evaluation --- has become a popular approach for tackling complex and nuanced questions across disciplines~\cite{heale2013understanding, tashakkori2007new}. For example, as discussed in~\cref{subsec:how}, using mixed methods allows researchers to blend qualitative and quantitative insights.

Triangulation is also important for establishing credibility, as no method is without its limitations. 
For example, user preferences and benchmark results may not correlate with user task performance, and different methods will help reveal a richer set of insights in combination~\cite{Mozannar2024realhumaneval}. 

In the current landscape of evaluation, extrinsic evaluations may overly rely on self-reported measures. Specifically, eight of the 25 papers that included a quantitative extrinsic evaluation only reported users' Likert scale ratings. While self-perceived ratings provide useful information about the system, they are also subject to cognitive biases and can be unreliable~\cite{elangovan2024considers,leung2011comparison,bishop2015use}. Self-reported results can be supplemented with other measurements. For example, \citet{kim_evallm:_2024} and \citet{liu_how_2024} analyze logs (e.g., click behavior, length of input) to understand how users interact with the system at a more granular level.

\paragraph{Ground methods in existing practices} As a product of triangulation, researchers may draw on methods outside their field. In this case, it is important to base techniques in existing practices. For example, seven papers from NLP venues include qualitative results (four offer case studies and three present select results as examples). One concern is that they do not justify how case studies were created or how qualitative examples were selected, raising concerns about methodological rigor. 

To demonstrate how we can ground our methods in other fields, we refer to HCI papers for examples on how to conduct qualitative analyses. For inductive analyses, papers often adopt a grounded-theory approach and discuss how they developed codes~\cite{thornberg2014grounded}. For example, \citet{arawjo_chainforge:_2024} analyzed interview data ``through a combination of inductive thematic analysis through affinity diagramming.'' For deductive analysis, \citet{lee_paperweaver:_2024} introduced a rubric with five dimensions. They provided details on dimensions that were created, defined, and then applied across annotators. See \cref{appen:qual_results} for examples of how qualitative methods are described.

\subsubsection{Rigorously evaluate evaluation}
It is essential to critically assess the evaluation methodologies themselves to understand the quality of findings and support replication in the future. 
Consistent with prior work~\cite{thomson2024common, card2020little}, we find there is a lack of rigorous meta-evaluations or validation of evaluation methodologies across our corpus. For example, the number of users participating in the evaluation tends to be small --- only five of the 39 papers we surveyed included more than $30$ participants in their sample size. This can lead to limited generalizability, introduce additional biases, and risk underpowering statistical tests~\cite{christley2010power}.

\paragraph{Expand meta-evaluation methods} Current practices for measuring reliability and validity are also constrained in scope. For reliability, papers focused almost exclusively on inter-rater reliability. Only one study reported on internal reliability (measured using Cronbach's $\alpha$). Other methods, such as test-retest or split-half, are not employed. For validity measures, papers mentioned using randomized controlled experiment designs, employing counterbalancing, and drawing from pre-validated surveys (e.g., System Usability Scale~\cite{bangor2008empirical}, NASA Task Load Index~\cite{hart2006nasa}). We suggest researchers consider other practices to ensure validity when applicable to customized evaluations, such as factor analysis, which is commonly used to check the validity of evaluation items in educational tests and surveys \cite{knekta2019one}. 

\paragraph{Document evaluation practices for replication} Meta-evaluation plays a critical part in stewarding scientific best practices. In theory, evaluation methods should be clearly documented and reproducible by others in the community; however, in practice, problems with replication have plagued both the NLP and HCI communities~\cite{belz2023missing, echtler2018open}. Our \evalname card aims to facilitate the design and documentation of evaluation to support replication.

\section{Case Study}

\label{sec:casestudy}
{Finally, we present two case studies demonstrating how \evalname can be used both for designing and reproducing evaluations.} 

\subsection{Using \evalname to design evaluations}
{We recruited two first authors (A1, A2) from the sample of 39 papers surveyed in \cref{sec:takeaways}. The authors were first asked to reflect on how they would improve the evaluation design from their paper. Then, they were asked to create a \evalname card for their paper before repeating the reflection process.}

{The authors reported that using \evalname encouraged them to engage in deeper considerations about the extrinsic implications of their systems and the validity of their evaluations. For example, after creating the \evalname card, both authors discussed integrating a long-term deployment study as part of their evaluation plan to \emph{``understand how the system can help users in their daily workflow''} (A1) and \emph{``to increase ecological validity''} (A2). A1 also discussed integrating statistical testing for repeated model evaluation, arising from the section on meta-evaluation in \evalname's framework.}

\subsection{Using \evalname to reproduce evaluations}
{\evalname serves not only as a design tool but also as documentation. In our second case study, we recruited two PhD students --- one who focuses on NLP (P1) and the other on HCI (P2). The participants were asked to write a reproduction plan for a human-AI system from the other domain using only the paper as reference. After writing the initial plan, they were given the \evalname card for the paper and asked to revise their plans. Then, we presented the reproduction plans written before seeing the \evalname card and after to the original authors of the paper to evaluate. }

{Participants reported improved understanding and confidence in reproducing the system's evaluation when given the \evalname card. As P1 stated, before having access to \evalname they found it difficult to understand the overarching design of the evaluation. In contrast, \evalname ``\emph{allow[ed] for quick understanding}'' (P2) of the evaluation plan. Moreover, plans created when participants had access to \evalname were more detailed and accurate. Before having access to \evalname, P2's reproduction plan only included the user study from the paper, overlooking the automated evaluation and case study. However, these details were included in the reproduction plan after having access to \evalname.}

\section{Conclusion}
We introduce the \evalname evaluation card
that can be used a priori to help researchers design more robust evaluations and post-hoc to standardize documentation on evaluation protocols. \evalname includes five key dimensions: what is being evaluated; how is evaluation conducted; who is participating in the evaluation; when is evaluation conducted; and how is the evaluation validated. Using \evalname, we survey 39 papers presenting new human-LLM systems published in HCI and NLP venues and present recommendations charting how evaluation practices should improve going forward. Through the adoption of \evalname evaluation card, we hope to facilitate new evaluation practices that are more realistic, rigorous, and reproducible.

\section*{Limitations}

This work has several limitations stemming from the scope and methodology of our review and framework application. 
First of all, we applied the \evalname card on a set of 39 *CL or CHI papers with human-AI systems. There is a vast space of venues that might publish on human-AI systems, such as domain-specific applications (e.g., AIED, medical journals) and trustworthy AI (e.g., FAccT). By selecting papers from only *CL and CHI, we may not fully capture the breadth of evaluation practices in diverse fields and miss some domain-specific insights. 
Additionally, our focus on human-AI systems that require an explicit interface may have excluded more NLP-centered studies that do not meet this criterion. 

Furthermore, we limited our inquiry on human-AI systems to human-LLM systems. 
While this focus is in response to the wide adoption of LLMs in current human-AI system research, it excludes other AI modalities, such as vision-centric systems, which may present unique evaluation challenges and opportunities. 
\evalname has the potential to be applied as a model-agnostic evaluation framework; nonetheless, it should be viewed as a starting point for broader inquiries into human-AI system evaluation. Future work could aim to include a more diverse and representative set of studies and explore evaluation practices across a wider range of AI capabilities and application areas.

{Finally, we evaluate \evalname using a case study that qualitatively examines how practitioners may apply the card across two settings. In future work, we encourage more comprehensive evaluation methods, such as quantitative analyses on how effective \evalname is, or qualitative studies with think-aloud protocols, similar to \citet{boyd2021datasheets}'s evaluation of Datasheets for Datasets~\cite{gebru2021datasheets}, to better understand practitioners' thought processes when using \evalname.}

\section*{Potential Risks}
A potential risk of this work is having documentation serve solely as an additional burden upon researchers, instead of fostering any positive change for human-AI system evaluation. For example, \citet{heger2022understanding} interviewed machine learning practitioners on using datasheets~\cite{gebru2021datasheets} for documenting datasets. They found that many of the practitioners they interviewed prioritized efficiency, and viewed documentation as taking time away from more important tasks. Some participants would complete the datasheets with the minimal amount of information required. In this case, \evalname runs the risk of being completed performatively, which deviates from our intentions of having the framework foster discussion about how to design evaluations and improve transparency for others in the community. 
\section*{Acknowledgment}
Dora Zhao is funded by the Brown Institute for Media Innovation. Xinran Zhao is supported by the ONR Award N000142312840.
We thank the OpenAI research access program for partial support of this work. We also thank anonymous reviewers for helpful discussions and comments.

\bibliographystyle{acl_natbib}
\bibliography{ref}

\appendix
\section{Methodology Details}
\label{appen:method}

\begin{figure*}[ht!]
    \centering
    \includegraphics[width=0.9\textwidth]{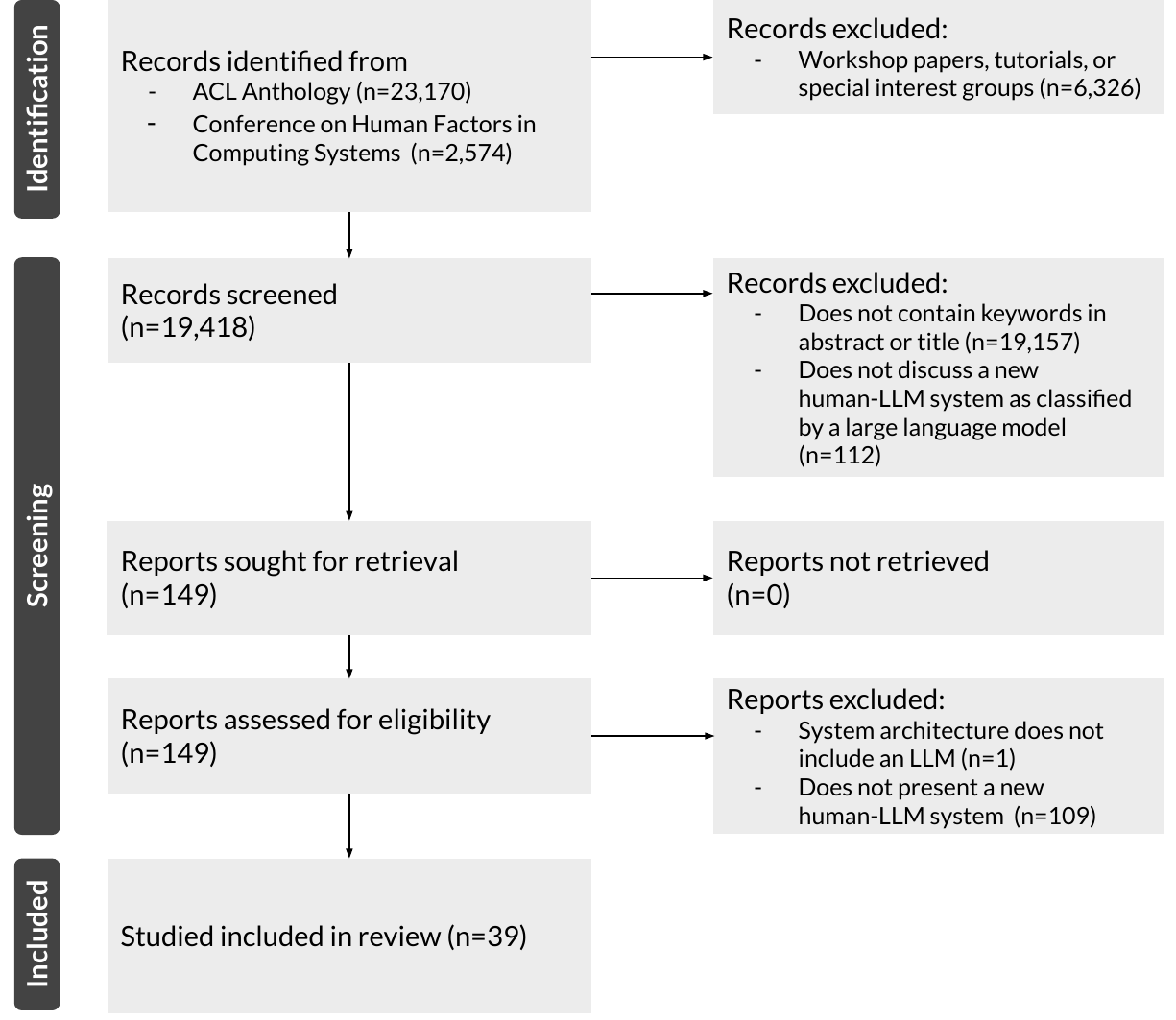}
    \caption{PRISMA diagram depicting the search strategy used to identify human-LLM systems for inclusion in our literature review.}
    \label{fig:filtering}
\end{figure*}
\subsection{Filtering Criteria}
We surveyed 39 papers introducing human-LLM systems which were identified using the process outlined in~\cref{fig:filtering}. First, we collected papers from venues focused on human-computer interactions (CHI) and natural language processing (*CL released on ACL Anthology). We filtered using regular expressions to find papers that contained keywords \texttt{ {[“human*”} OR “user*”] AND ``system*'' AND [``large language model*'' OR ``LLM*'']}. We also used a large language model to filter whether the abstract and title discussed a new human-LLM system using the following prompt: \\\\
\texttt{``You are a helpful literature review assistant whose job is to read the below paper and help me decide if it satisfies all my criteria. The criteria are:
\\
(1) The paper presents a human-LLM interaction system and evaluate the system in some ways.\\
(2) The system described in the paper must have human interact with large language model in some ways.\\
The paper is:\\
Title: \${TITLE}\\
Abstract: \${ABSTRACT}\\\\
Please give me a binary answer (yes/no) on whether this paper satisfies all the criteria (you should say no as long as any one of the criteria is not met). Do not include any other explanation, the output should be either yes or no.''}

\subsection{Papers Reviewed}
The outcome of the annotation using the taxonomy in \cref{tab:taxonomy} and the coding guide in \cref{appen:codes} is presented in \cref{fig:labels}.

\subsubsection{Papers published at an HCI Venue}
\citet{lee_paperweaver:_2024}; \citet{lawley_val:_2024}; \citet{liu_selenite:_2024}; \citet{zavolokina_think_2024}; \citet{cheng_relic:_2024}; \citet{calle_towards_2024}; \citet{yu_reducing_2024}; \citet{zulfikar_memoro:_2024}; \citet{liu_human_2024}; \citet{taeb_axnav:_2024}; \citet{wu_ai_2022}; \citet{wang_virtuwander:_2024}; \citet{rajashekar_human-algorithmic_2024}; \citet{arawjo_chainforge:_2024}; \citet{wang_popblends:_2023}; \citet{zhang_see_2024}; \citet{kim_evallm:_2024}; \citet{fan_contextcam:_2024}; \citet{liu_how_2024}; \citet{liu_what_2023}
\subsubsection{Papers published at an NLP Venue}
\citet{zhao_narrativeplay:_2024}; \citet{ding_harnessing_2023}; \citet{ma_insightpilot:_2023}; \citet{cai_low-code_2024}; \citet{chakrabarty_help_2022}; \citet{gloria-silva_plan-grounded_2024}; \citet{fei_empathyear:_2024}; \citet{raheja_coedit:_2023}; \citet{inan_generating_2024}; \citet{wei_collabkg:_2024}; \citet{addlesee_multi-party_2024}; \citet{liu2024proofread}; \citet{yang_beyond_2023}; \citet{kim_meganno+:_2024}; \citet{hu_dialight:_2024}; \citet{luo_duetsim:_2024}; \citet{ni_chatreport:_2023}; \citet{navarro_exploring_2023}

\begin{figure*}[!htbp]
    \centering
    \includegraphics[width=\linewidth]{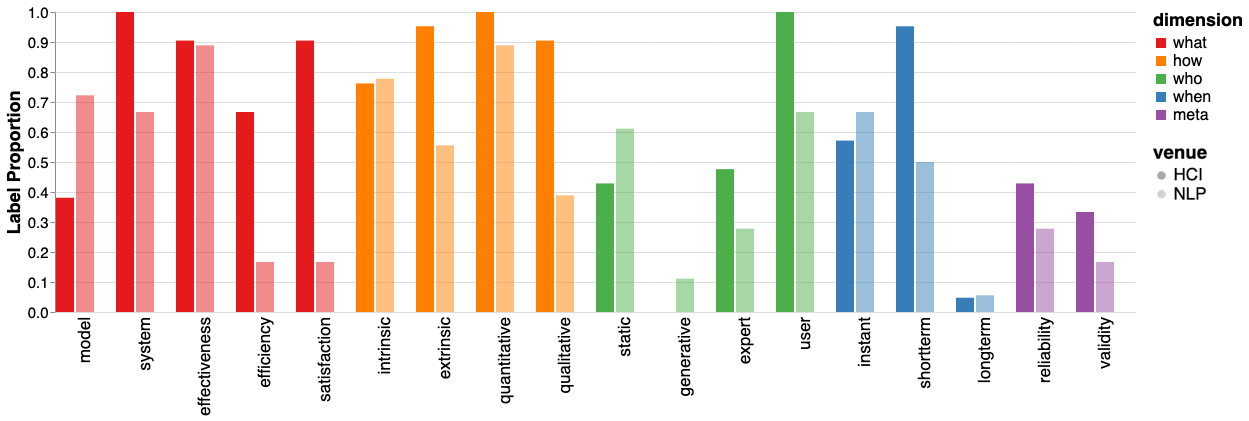}
    \vspace{-15pt}
    \caption{Distribution of evaluation annotations on the 39 papers by HCI or NLP venues using \evalname.}
    \label{fig:labels}
\end{figure*}

\section{Annotation Guide}
\label{appen:codes}
We provide the codebook used to label the evaluations conducted in each paper below. This codebook is also available in our \href{https://github.com/sphere-eval/sphere-eval.github.io/blob/main/data/annotation_guide.md}{GitHub repository}.

\setlist[itemize,enumerate]{leftmargin=10pt}

\subsection{What is being evaluated?}
\subsubsection{Component types}
\begin{enumerate}
    \item \textbf{what\_model}: The evaluation focuses on only the model capabilities, including the model's performance pre-deployment in traditional benchmarking settings and performance in-situ as users continue to interact with the model. Mark 1 if the authors evaluate the model and 0 if not.
    \begin{itemize}
        \item For example: The authors benchmark the performance of a fine-tuned model used in their system.
    \end{itemize}
    
    \item \textbf{what\_system}: The evaluation covers the system as a whole to understand how these design choices may impact users' experiences, including different layers of the system, such as interfaces or interactions. Mark 1 if the authors evaluate the system and 0 if not.
\end{enumerate}

\subsubsection{Design goals}
For each design goal, mark 1 if the authors include evaluation that covers the concept and 0 otherwise.

\begin{enumerate}
    \item \textbf{what\_effectiveness}: Evaluating the accuracy, completeness, and lack of negative consequences with which users achieved specified goals.
    \begin{itemize}
        \item For example: Are users able to successfully complete the task the system is designed for? What is the performance of the model?
    \end{itemize}

    \item \textbf{what\_efficiency}: Evaluating resources (such as time or effort) needed by users to achieve their goals.
    \begin{itemize}
        \item For example: How long did it take for a user to complete the task? What was the cognitive burden or mental load required to complete the task?
    \end{itemize}

    \item \textbf{what\_satisfaction}: Evaluating positive attitudes, emotions, and/or comfort resulting from use of a system, product, or service.
    \begin{itemize}
        \item For example: What was the user’s satisfaction with the overall system or parts of the system? How much did the user trust the system?
    \end{itemize}
\end{enumerate}

\subsection{How is the evaluation being conducted?}

\subsubsection{Scope}
\begin{enumerate}
    \item \textbf{how\_intrinsic}: Assessing the system or the internal model components on specific tasks that they are designed to perform, by evaluating how well they achieve these tasks according to some predefined criteria or benchmarks. Mark 1 if they include intrinsic evaluation and 0 if not.
    \begin{itemize}
        \item For example, in a writing assistant system, intrinsic evaluation might involve assessing the system's accuracy on grammatical correctness with automatic metrics or how users might rate the functionality of different system features.
    \end{itemize}

    \item \textbf{how\_extrinsic}: Measuring the effectiveness of the system in the context of its application in real-world scenarios, when interacting with users. Mark 1 if they include extrinsic evaluation and 0 if not.
    \begin{itemize}
        \item For example, in a writing assistant system, extrinsic evaluation might involve recruiting users to co-write with the system and seeing how this impacts their writing style, productivity, and satisfaction.
    \end{itemize}
\end{enumerate}

\subsubsection{Method}
\begin{enumerate}
    \item \textbf{how\_quantitative}: Measuring and analyzing numerical data to assess system performance and impact. Examples include measuring Likert scale ratings or running benchmark evaluations on the model. Mark 1 if the authors included any quantitative methods and 0 if not.

    \item \textbf{how\_qualitative}: Analyzing non-numerical data to gain deeper insights into user experiences, perceptions, and the contextual factors influencing system performance. Examples include conducting semi-structured interviews with users. Mark 1 if the authors included any qualitative methods and 0 if not.
    \begin{itemize}
        \item For example, in NarrativePlay, the authors include some brief analysis of responses. However, we do \textit{not} count listing examples as qualitative analysis.
    \end{itemize}
\end{enumerate}

\subsection{Who is participating in the evaluation process?}

\subsubsection{Automated Evaluators}
Use these tags \textit{only} when evaluation not involving human participants is used.

\begin{enumerate}
    \item \textbf{who\_static}: Mark 1 if any static evaluation not directly performed by a human or LLM is included and 0 if not. For example, benchmarking a model's capability is an example of static evaluation.

    \item \textbf{who\_generative}: Mark 1 if the authors use a language model in a generative capacity and 0 if not for evaluation. Examples include simulating participants with LLMs, using LLM to annotate and rate text, or using LLM-as-a-judge. Using a technique like BERTScore is \textit{not} generative since it is embedding-based.
\end{enumerate}

\subsubsection{Human Evaluators}
Use these tags \textit{only} when human evaluation is used.

\begin{enumerate}
    \item \textbf{who\_expert}: If the human evaluator is a domain expert or has equivalent expertise in the area that the system is designed for. Mark 1 if expert evaluators are included and 0 otherwise.
    \begin{itemize}
        \item For example, if the system is a tutoring system, a teacher would be considered an expert.
        \item In AngleKindling, which helps journalists come up with framings for papers, the evaluation is conducted with NYC journalists, who are domain experts in this field.
    \end{itemize}

    \item \textbf{who\_user}: If the evaluator is a direct user or target audience of the system. Mark 1 if the intended user is included as a human evaluator and 0 otherwise.
    \begin{itemize}
        \item For example, for a student-facing tutoring system, student evaluators will be the design target, but a teacher will not.
        \item In AngleKindling, the journalists are also the intended users for the system, so we would mark 1.
        \item If you have a general-purpose system, any user (including crowdworkers, PhD students) would be a user.
    \end{itemize}
\end{enumerate}

\subsection{When is evaluation conducted?}
The time-scale over which the evaluation occurs.

\begin{enumerate}
    \item \textbf{when\_immediate}: Evaluating real-time or immediate interactions. Mark 1 if any immediate evaluation is conducted and 0 otherwise.

    \item \textbf{when\_shortterm}: Evaluation is deployed for a short duration of time to measure short-term benefits of using the system. Mark 1 if any short-term evaluation is conducted and 0 otherwise.

    \item \textbf{when\_longterm}: Evaluation is conducted over a longer duration of time. This is typically done to understand long-term behavioral changes, practical feasibility, etc. Mark 1 if any long-term evaluation is conducted and 0 otherwise.
\end{enumerate}

\subsection{How is evaluation validated?}
The methods for ensuring the reliability and validity of the evaluation methods. Mark 1 only if the authors \textit{explicitly} mention any techniques for reliability and validity.

\begin{enumerate}
    \item \textbf{reliability}: Mark 1 if the authors include techniques or measures to ensure that the evaluation judgments are consistent.
    \begin{itemize}
        \item For example, looking at internal consistency (e.g., inter-rater reliability, Cronbach alpha, split-half reliability), consistency over time (test-retest reliability), and reproducibility of results.
    \end{itemize}

    \item \textbf{validity}: Mark 1 if the authors include techniques or measures to ensure that the evaluation method measures the correct constructs.
    \begin{itemize}
        \item For example, methods include removing human biases using experiment designs, using statistical methods like factor analysis, mentioning ecological validity in their experiment setup, etc.
    \end{itemize}
\end{enumerate}

\section{Additional Results}
\subsection{Paper Annotations}
We provide the annotations for the 39 papers we reviewed using the \evalname framework in \cref{tab:anno_1} and \cref{tab:anno_2}.
The annotations are also available on our \href{https://sphere-eval.github.io/#annotated-papers}{website} and \href{https://github.com/sphere-eval/sphere-eval.github.io/blob/main/data/annotated_papers.csv}{GitHub repository}.

\begin{table*}[]
    \footnotesize
    \centering
    \begin{tabular}{lrrrrrrrrr}
    System & \rotatebox{45}{Model} & \rotatebox{45}{System} & \rotatebox{45}{Effective} & \rotatebox{45}{Efficiency} & \rotatebox{45}{Satisfac.} & \rotatebox{45}{Intrinsic} & \rotatebox{45}{Extrinsic} & \rotatebox{45}{Quant.} & \rotatebox{45}{Qual.}\\
    \midrule 
    \citet{addlesee_multi-party_2024} & \checkmark & \checkmark & \checkmark &  &  & \checkmark &  & \checkmark & \\
    \citet{calle_towards_2024} &  & \checkmark & \checkmark & \checkmark &  & \checkmark &  & \checkmark & \\
    \citet{ding_harnessing_2023} &  & \checkmark & \checkmark & \checkmark & \checkmark & \checkmark & \checkmark & \checkmark & \checkmark\\
    \citet{inan_generating_2024} & \checkmark & \checkmark & \checkmark &  & \checkmark & \checkmark & \checkmark & \checkmark & \\
    \citet{liu_what_2023} &  & \checkmark & \checkmark & \checkmark & \checkmark & \checkmark & \checkmark & \checkmark & \checkmark\\
    \citet{navarro_exploring_2023} & \checkmark &  & \checkmark &  &  & \checkmark &  & \checkmark & \\
    \citet{rajashekar_human-algorithmic_2024} &  & \checkmark &  & \checkmark & \checkmark &  & \checkmark & \checkmark & \checkmark\\
    \citet{wu_ai_2022} &  & \checkmark & \checkmark & \checkmark & \checkmark &  & \checkmark & \checkmark & \checkmark\\
    \citet{zhang_see_2024} &  & \checkmark & \checkmark &  & \checkmark &  & \checkmark & \checkmark & \checkmark\\
    AXNav~\cite{taeb_axnav:_2024} &  & \checkmark & \checkmark &  & \checkmark & \checkmark & \checkmark & \checkmark & \checkmark\\
    ChainForge~\cite{arawjo_chainforge:_2024} &  & \checkmark & \checkmark &  & \checkmark &  & \checkmark & \checkmark & \checkmark\\
    ChatReport~\cite{ni_chatreport:_2023} & \checkmark &  & \checkmark &  &  & \checkmark &  & \checkmark & \checkmark\\
    ClarifAI~\cite{zavolokina_think_2024} &  & \checkmark & \checkmark & \checkmark & \checkmark & \checkmark & \checkmark & \checkmark & \checkmark\\
    CoEdit~\cite{raheja_coedit:_2023} & \checkmark & \checkmark &  &  &  & \checkmark &  & \checkmark & \\
    CollabKG~\cite{wei_collabkg:_2024} &  & \checkmark & \checkmark & \checkmark &  &  & \checkmark & \checkmark & \\
    ContextCam~\cite{fan_contextcam:_2024} &  & \checkmark & \checkmark &  & \checkmark & \checkmark & \checkmark & \checkmark & \checkmark\\
    CoPoet~\cite{chakrabarty_help_2022} & \checkmark & \checkmark & \checkmark &  &  & \checkmark & \checkmark & \checkmark & \\
    CoQuest~\cite{liu_how_2024} &  & \checkmark & \checkmark & \checkmark & \checkmark & \checkmark & \checkmark & \checkmark & \checkmark\\
    Dialight~\cite{hu_dialight:_2024} & \checkmark &  & \checkmark &  &  & \checkmark & \checkmark & \checkmark & \\
    DuetSim~\cite{luo_duetsim:_2024} & \checkmark &  & \checkmark &  &  & \checkmark & \checkmark & \checkmark & \checkmark\\
    EmpathyEar~\cite{fei_empathyear:_2024} & \checkmark & \checkmark &  &  &  & \checkmark &  & \checkmark & \\
    EvalLM~\cite{kim_evallm:_2024} & \checkmark & \checkmark & \checkmark & \checkmark & \checkmark & \checkmark & \checkmark & \checkmark & \checkmark\\
    Human I/O~\cite{liu_human_2024} &  & \checkmark & \checkmark & \checkmark & \checkmark & \checkmark & \checkmark & \checkmark & \checkmark\\
    InsightPilot~\cite{ma_insightpilot:_2023} &  & \checkmark & \checkmark &  &  &  & \checkmark & \checkmark & \checkmark\\
    Low-code LLM~\cite{cai_low-code_2024} &  & \checkmark & \checkmark &  &  &  & \checkmark &  & \checkmark\\
    MEGAnno+~\cite{kim_meganno+:_2024} &  & \checkmark & \checkmark &  &  &  & \checkmark &  & \checkmark\\
    Memoro~\cite{zulfikar_memoro:_2024} & \checkmark & \checkmark & \checkmark & \checkmark & \checkmark & \checkmark & \checkmark & \checkmark & \checkmark\\
    NarrativePlay~\cite{zhao_narrativeplay:_2024} & \checkmark &  & \checkmark &  &  & \checkmark &  & \checkmark & \\
    NavNudge~\cite{yu_reducing_2024} &  & \checkmark & \checkmark & \checkmark & \checkmark &  & \checkmark & \checkmark & \\
    PaperWeaver~\cite{lee_paperweaver:_2024} & \checkmark & \checkmark & \checkmark & \checkmark & \checkmark & \checkmark & \checkmark & \checkmark & \checkmark\\
    PlanLLM~\cite{gloria-silva_plan-grounded_2024} & \checkmark & \checkmark & \checkmark &  &  & \checkmark &  & \checkmark & \\
    PopBlends~\cite{wang_popblends:_2023} & \checkmark & \checkmark & \checkmark & \checkmark & \checkmark & \checkmark & \checkmark & \checkmark & \checkmark\\
    Proofread~\cite{liu2024proofread} & \checkmark &  & \checkmark &  &  & \checkmark &  & \checkmark & \\
    Rehearsal~\cite{shaikh_rehearsal:_2024} & \checkmark & \checkmark & \checkmark &  &  & \checkmark & \checkmark & \checkmark & \checkmark\\
    RELIC~\cite{cheng_relic:_2024} & \checkmark & \checkmark & \checkmark & \checkmark & \checkmark & \checkmark & \checkmark & \checkmark & \checkmark\\
    Selenite~\cite{liu_selenite:_2024} & \checkmark & \checkmark & \checkmark & \checkmark & \checkmark & \checkmark & \checkmark & \checkmark & \checkmark\\
    VAL~\cite{lawley_val:_2024} & \checkmark & \checkmark & \checkmark &  & \checkmark & \checkmark & \checkmark & \checkmark & \checkmark\\
    VirtuWander~\cite{wang_virtuwander:_2024} &  & \checkmark &  &  & \checkmark & \checkmark & \checkmark & \checkmark & \checkmark\\
    Weaver~\cite{yang_beyond_2023} & \checkmark & \checkmark & \checkmark & \checkmark & \checkmark & \checkmark & \checkmark & \checkmark & \checkmark\\
    \bottomrule
    \end{tabular}
    \caption{Annotations on the 39 human-LLM systems covering the dimensions of what is being evaluated and how evaluation is conducted.}
    \label{tab:anno_1}
\end{table*}
\begin{table*}[]
    \footnotesize
    \centering
    \begin{tabular}{lrrrrrrrrr}
    System & \rotatebox{45}{User} & \rotatebox{45}{Expert} & \rotatebox{45}{Static} & \rotatebox{45}{Generative} & \rotatebox{45}{Immediate} &\rotatebox{45}{Short} & \rotatebox{45}{Long} & \rotatebox{45}{Reliable} & \rotatebox{45}{Valid}\\
    \midrule
    \citet{addlesee_multi-party_2024} & \checkmark &  &  &  & \checkmark &  &  &  & \\
    \citet{calle_towards_2024} & \checkmark &  & \checkmark & \checkmark & \checkmark &  &  &  & \\
    \citet{ding_harnessing_2023} &  & \checkmark &  &  & \checkmark & \checkmark &  &  & \checkmark\\
    \citet{inan_generating_2024} & \checkmark &  & \checkmark & \checkmark & \checkmark & \checkmark & \checkmark &  & \\
    \citet{liu_what_2023} & \checkmark & \checkmark &  &  &  & \checkmark &  & \checkmark & \\
    \citet{navarro_exploring_2023} & \checkmark &  &  &  & \checkmark &  &  &  & \\
    \citet{rajashekar_human-algorithmic_2024} & \checkmark & \checkmark &  &  & \checkmark & \checkmark &  &  & \checkmark\\
    \citet{wu_ai_2022} & \checkmark & \checkmark &  &  & \checkmark & \checkmark &  &  & \\
    \citet{zhang_see_2024} &  & \checkmark &  &  &  & \checkmark &  &  & \\
    AXNav~\cite{taeb_axnav:_2024} & \checkmark & \checkmark & \checkmark &  & \checkmark & \checkmark &  & \checkmark & \\
    ChainForge~\cite{arawjo_chainforge:_2024} &  & \checkmark &  &  &  & \checkmark &  &  & \\
    ChatReport~\cite{ni_chatreport:_2023} & \checkmark &  &  &  &  &  &  & \checkmark & \\
    ClarifAI~\cite{zavolokina_think_2024} & \checkmark & \checkmark &  &  &  & \checkmark &  & \checkmark & \\
    CoEdit~\cite{raheja_coedit:_2023} & \checkmark &  & \checkmark &  & \checkmark &  &  &  & \\
    CollabKG~\cite{wei_collabkg:_2024} &  &  &  & \checkmark &  & \checkmark &  &  & \\
    ContextCam~\cite{fan_contextcam:_2024} &  & \checkmark & \checkmark &  & \checkmark & \checkmark & \checkmark &  & \\
    CoPoet~\cite{chakrabarty_help_2022} & \checkmark &  &  & \checkmark & \checkmark & \checkmark &  & \checkmark & \checkmark\\
    CoQuest~\cite{liu_how_2024} & \checkmark & \checkmark &  &  &  & \checkmark &  & \checkmark & \\
    Dialight~\cite{hu_dialight:_2024} & \checkmark &  &  & \checkmark & \checkmark &  &  &  & \\
    DuetSim~\cite{luo_duetsim:_2024} &  & \checkmark & \checkmark &  & \checkmark &  &  &  & \\
    EmpathyEar~\cite{fei_empathyear:_2024} & \checkmark &  &  & \checkmark & \checkmark &  &  &  & \\
    EvalLM~\cite{kim_evallm:_2024} & \checkmark & \checkmark & \checkmark &  & \checkmark & \checkmark &  & \checkmark & \\
    Human I/O~\cite{liu_human_2024} &  & \checkmark & \checkmark &  & \checkmark & \checkmark &  & \checkmark & \checkmark\\
    InsightPilot~\cite{ma_insightpilot:_2023} &  & \checkmark &  &  &  & \checkmark &  &  & \\
    Low-code LLM~\cite{cai_low-code_2024} &  &  &  &  &  &  &  &  & \\
    MEGAnno+~\cite{kim_meganno+:_2024} & \checkmark & \checkmark &  &  &  & \checkmark &  &  & \\
    Memoro~\cite{zulfikar_memoro:_2024} &  & \checkmark &  &  & \checkmark & \checkmark &  &  & \checkmark\\
    NarrativePlay~\cite{zhao_narrativeplay:_2024} &  & \checkmark &  & \checkmark & \checkmark & \checkmark &  & \checkmark & \\
    NavNudge~\cite{yu_reducing_2024} &  &  &  & \checkmark &  & \checkmark &  &  & \checkmark\\
    PaperWeaver~\cite{lee_paperweaver:_2024} & \checkmark & \checkmark &  &  &  & \checkmark &  & \checkmark & \checkmark\\
    PlanLLM~\cite{gloria-silva_plan-grounded_2024} & \checkmark & \checkmark &  & \checkmark & \checkmark & \checkmark &  & \checkmark & \\
    PopBlends~\cite{wang_popblends:_2023} & \checkmark & \checkmark &  &  &  & \checkmark &  & \checkmark & \\
    Proofread~\cite{liu2024proofread} & \checkmark &  &  &  &  &  &  &  & \\
    Rehearsal~\cite{shaikh_rehearsal:_2024} &  & \checkmark & \checkmark &  & \checkmark & \checkmark &  & \checkmark & \\
    RELIC~\cite{cheng_relic:_2024} &  & \checkmark & \checkmark &  & \checkmark & \checkmark &  &  & \\
    Selenite~\cite{liu_selenite:_2024} &  & \checkmark & \checkmark &  & \checkmark & \checkmark &  &  & \checkmark\\
    VAL~\cite{lawley_val:_2024} &  & \checkmark & \checkmark &  & \checkmark & \checkmark &  &  & \\
    VirtuWander~\cite{wang_virtuwander:_2024} &  & \checkmark &  &  &  & \checkmark &  &  & \\
    Weaver~\cite{yang_beyond_2023} & \checkmark & \checkmark & \checkmark &  & \checkmark & \checkmark &  & \checkmark & \checkmark\\
    \bottomrule
    \end{tabular}
    \caption{Annotations on the 39 human-LLM systems covering the dimensions of who is participating in the evaluation, when is the evaluation (duration), and how evaluation is evaluated.}
    \label{tab:anno_2}
\end{table*}
\subsection{Quotations}
\label{appen:qual_results}
We provide quotations from papers in our literature review describing their qualitative methods in Table~\ref{tab:qual}.

\begin{table*}[hb]
    \small
    \centering
    \begin{tabular}{p{0.95\linewidth}}
    \toprule
     ``We analyzed the transcripts through a combination of inductive thematic analysis through affinity diagramming, augmented with a spreadsheet to list participants’ ideas, behaviors (nodes added, process of their exploration, whether they imported data, etc), and answers to post-interview questions. For our in-lab study, three coauthors separately affinity diagrammed three transcripts each, then met and joined the clusters through mutual discussion. The merged cluster was iteratively expanded with more participant data until clusters reached saturation. For interviews, the first author affinity diagrammed all transcripts to determine themes.''--~\citet{arawjo_chainforge:_2024} \\
     \midrule
     ``The research team also took field notes during the session and used the notes to guide the analysis...We performed a thematic analysis on the qualitative data from the user study. Two authors of the paper first individually coded all the transcripts, then presented the codes to each other and collaboratively and iteratively constructed an affinity diagram of quotes and codes together to develop themes.''--~\citet{taeb_axnav:_2024}\\
     \midrule
     ``Think-aloud data was primarily used for understanding how users generated and interpreted RQs. One researcher first generated a codebook through open coding using videos and transcripts from three randomly selected participants, and then three other researchers independently coded the data of the same three participants, reaching an inter-rater agreement of 0.83 in Krippendorf’s alpha. The annotators then discussed and refined the codebook again until they reached full agreement. Then, four researchers proceeded to annotate the remaining 17 participants’ behavior data separately. In the final codebook, whether users interacted with the system was annotated and used for quantitative analysis in RQ3 as “Acted During Wait”. The final codebook also included sense-making behavior (e.g., reasons for (not) waiting, reason for providing certain feedback) as qualitative results.''--~\citet{liu_how_2024}\\
     \midrule
     ``All study sessions were recorded and transcribed. Two authors read through the text script of three randomly selected participants together to understand their user experience of the prototype. Then, they independently coded the script using an open-coding approach. They combined deductive and inductive coding techniques to form the codebook. The two coders regularly discussed the codes and resolved disagreements to create a consolidated codebook. Further meetings were scheduled with the whole research team to discuss the codes and how they should be grouped into themes. The whole team iterated on the codes and their grouping until they reached consensus. In the end, we arrived at four themes: overall user behavioral patterns, engagement, diverse information, and in-depth information processing''--~\citet{zhang_see_2024}\\
     \bottomrule
    \end{tabular}
    \caption{Set of examples of how qualitative methods were described in papers from HCI venues.}
    \label{tab:qual}
\end{table*}

\section{Case Studies}
\label{appen:case_study}
We present examples of evaluation cards for two human-AI systems: LearnLM~\cite{Jurenka2024learnLM} (\cref{fig:learnlm-card}) and AngleKindling~\cite{petridis2023anglekindling} (\cref{fig:angle-card}).
We selected these two systems as examples since they span different application domains (education and journalism) and different modes of model development commonly observed in current human-AI systems, ranging from fine-tuning and aligning LLM (LearnLM) to prompt engineering (AngleKindling).

\subsection{Takeaways}
Using \evalname, in addition to describing existing system evaluations, we can also identify areas of improvement in evaluation practices for our case studies: extrinsic measures, long-term evaluation, and validations.

First, both cases provide both intrinsic and extrinsic evaluations, but there is still a lack of \textit{quantitative extrinsic} evaluation for the overall \textit{effectiveness} and \textit{efficiency} of the \textit{system}. The current quantitative extrinsic evaluation focuses on self-perceived ratings of the systems' performance. While capturing users' perceptions is important, self-perceived ratings may be unreliable and subject to human bias. Including objective, quantitative measures can provide a more holistic picture of system performance. For example, in LearnLM, experts provided ratings for each turn generated during a tutoring session. We could also measure changes in student performance using pre-post test scores, grades, or drop-out rates. 

Second, both studies only perform immediate or short-term evaluations. Thus, we do not know the downstream impact that these systems might have. For example, for AngleKindling, a longitudinal evaluation could help us understand how journalists integrate the system into their workflow and the potential impact on productivity or the types of stories being written. 

Finally, while there is some acknowledgment of meta-evaluation done in the case of LearnLM, overall discussion on validating the evaluation paradigm is not common.

\begin{figure*}

\centering
        \begin{tcolorbox}[
colback=orange!5!white,colframe=orange!35!white,title=SPHERE Evaluation Card: LearnLM-Tutor]
\small
\emph{\textbf{System Overview:} LearnLM-Tutor \cite{Jurenka2024learnLM} is conversational AI designed to serve as a personalized tutor for learners and a teaching assistant for educators. The model component is a fine-tuned Gemini 1.0~\cite{team2023gemini} using a custom dataset. LearnLM-Tutor is evaluated using seven different methods of which we only present two here as examples: turn-level pedagogy (TLP, \href{https://goo.gle/LearnLM}{\S 5.2}), and language model evaluation (LME, \href{https://goo.gle/LearnLM}{\S 6.1}).} \\

\textbf{What is being evaluated?}
\begin{itemize}[noitemsep,topsep=0pt] 
    \item \textbf{Component:} TLP \& LME both evaluated the \textit{model} component of LearnLM-Tutor. 
    \item \textbf{Design Goal:} Evaluate LearnLM-Tutor’s \emph{effectiveness} (TLP: how good is each turn's pedagogy use, and LME: how well it performs on different pedagogical tasks).
\end{itemize} \vspace{5pt}
\textbf{How is evaluation conducted? }
\begin{itemize}[noitemsep,topsep=0pt]
    \item \textbf{Scope:} For both TLP \& LME, an \emph{intrinsic} evaluation is conducted to examine the LearnLM model's capacity. 
    \item \textbf{Method:} 
    For TLP, each turn of unguided tutoring sessions between real learners and either LearnLM-Tutor or Gemini 1.0 was rated on yes/no/na for nine items of customized pedagogy rubrics (e.g., promotes engagement, monitors motivation, etc.). Welch t-test with Holm-Bonferroni adjustment was used to compare the turn-level \textit{quantitative} scores of LearnLM-Tutor and Gemini 1.0. \\
    For LME, they make the LLM critic generate \textit{quantitative} binary scores on an expert-curated dataset of different customized pedagogical tasks (e.g., stay on topic, don't reveal the answer, etc.), using task-specific prompts (including task description, reference answer, context, and the generated response).
\end{itemize} \vspace{5pt}
\textbf{Who is participating in the evaluation? }
\begin{itemize}[noitemsep,topsep=0pt]
    \item \textbf{Automated:} PaLM 2.0 is used as a \textit{generative} evaluator for the model for LME. 
    \item \textbf{Human:} Approximately 60 human pedagogical \textit{experts} were recruited to give ratings for TLP (\textit{intended users} for LearnLM-Tutor would be the students). There were about 1.5k overall turns for each model rated by at least three different raters, and a majority vote was used for each model's response. 
\end{itemize} \vspace{5pt}
\textbf{When is evaluation conducted? }
\begin{itemize}[noitemsep,topsep=0pt]
    \item \textbf{Time Scale}: LME is conducted \textit{immediately} since it uses a \textit{generative} evaluator. For TLP, there is no explicit mention of the time it takes to label each turn for a human expert, so we mark it as \emph{immediate} as well.
\end{itemize} \vspace{5pt}
\textbf{How is evaluation validated?}
\begin{itemize}[noitemsep,topsep=0pt]
    \item \textbf{Validation}: For TLP, the Krippendorff's $\alpha$ of 0.3 across all attributes showed a low \textit{inter-rater reliability}. For LME, the paper reported that the average LME score highly correlates with humans for different iterations of the tutor model. However, in terms of \textit{validity}, the paper lacks details on the score calculation, correlation statistics, and how the pedagogy rubrics for LME were developed.
\end{itemize}
\end{tcolorbox}

\footnotesize
\begin{tcbitemize}[raster columns=2,raster equal height, colframe=yellow!55!red,colback=orange!5!white,fonttitle=\bfseries, subtitle style={boxrule=0pt,
colback=yellow!50!red!25!white,
colupper=orange!75!gray} ]

\tcbitem[squeezed title*={Turn-Level Pedagogy (TLP, \href{https://goo.gle/LearnLM}{\S 5.2})}]
\tcbsubtitle{What is being evaluated?}
\textit{Model} component’s \emph{effectiveness}: how good is each turn's pedagogy use.

\tcbsubtitle{How is evaluation conducted? }
\emph{Intrinsic} evaluation: each turn of unguided tutoring sessions was rated on yes/no/na for nine rubric items (e.g., promotes engagement, monitors motivation, etc.). Welch t-test with Holm-Bonferroni adjustment was used to compare the turn-level \textit{quantitative} scores of LearnLM-Tutor and Gemini 1.0. 

\tcbsubtitle{Who is participating in the evaluation? }
Approximately 60 pedagogical \textit{experts}. About 1.5k overall turns rated by at least three different raters, and a majority vote was used for each model's response. 

\tcbsubtitle{When is evaluation conducted? }
No explicit mention of time it takes to label each turn for a human expert, so we mark it as \emph{immediate}.
 
\tcbsubtitle{How is evaluation validated?}
The Krippendorff's $\alpha$ of 0.3 across all attributes showed a low \textit{inter-rater reliability}

\tcbitem[squeezed title*={Language Model Evaluation (LME, \href{https://goo.gle/LearnLM}{\S 6.1})}]
\tcbsubtitle{What is being evaluated?}
\textit{Model} component’s \emph{effectiveness}: how well it performs on different pedagogical tasks.

\tcbsubtitle{How is evaluation conducted? }
\emph{Intrinsic} evaluation: LLM critic generates \textit{quantitative} binary scores on an expert-curated dataset of different customized pedagogical tasks (e.g., stay on topic, don't reveal the answer, etc.), using task-specific prompts (including task description, reference answer, context, and the generated response).

\tcbsubtitle{Who is participating in the evaluation? }
PaLM 2.0 is used as a \textit{generative} evaluator for the model for LME. 

\tcbsubtitle{When is evaluation conducted? }
\textit{Immediate} since it uses a \textit{generative} evaluator.

\tcbsubtitle{How is evaluation validated?}
Average LME score highly correlates with humans for different iterations of the tutor model. However, the paper lacks details on the \textit{validity} for score calculation, correlation statistics and rubrics development.

\end{tcbitemize}
    
    \caption{Example \evalname evaluation card for LearnLM-Tutor~\cite{Jurenka2024learnLM}. One can apply \evalname with one card per human-AI system as in \cref{fig:angle-card}, or one card per evaluation method for cleaner separation.}
    \label{fig:learnlm-card}
\end{figure*}

\begin{figure*}
    \centering 
    \begin{tcolorbox}[
colback=blue!5!white,colframe=blue!35!white,title=SPHERE Evaluation Card: AngleKindling]
\small
\emph{\textbf{System Overview:} AngleKindling is designed to help journalists brainstorm different ideas for stories from press releases. The model component includes few-shot prompting on \texttt{GPT-3} to extract the main points of press releases and propose different angles. The user interface displays generated results linked to previous New York Times articles and historical background.} \\

\textbf{What is being evaluated?}
\begin{itemize}
    \item \textbf{Component:} The authors conducted an evaluation of their \emph{system} component.
    \item \textbf{Design Goal:} AngleKindling’s \emph{effectiveness} (how many pursuable angles were created), \emph{efficiency} (mental demand), and \emph{user satisfaction} (how much they liked different features and overall helpfulness) were evaluated. 
\end{itemize}
\textbf{How is evaluation conducted? }
\begin{itemize}
    \item \textbf{Scope:} An \emph{extrinsic} evaluation is conducted. 
    \item \textbf{Method:} Participants were first interviewed about their journalism background. Then, in the {within-subjects user study}, participants were asked to use both AngleKindling and INJECT, an existing support tool for journalists, to brainstorm angles for a press release. They then answered a questionnaire after using each tool. Finally, they participated in a \emph{semi-structured interview (qualitative)}.  
    
    The questionnaire had participants rate the following dimensions on a 7-point Likert scale \emph{(quantitative)}: helpfulness, pursuable angles, and mental demand. Participants also rated how helpful individual features were on both AngleKindling and INJECT. Paired-sample Wilcoxon tests with Bonferroni correction between the Likert scale ratings of Helpfulness, Pursuable Angles, and Mental Demand for AngleKindling versus INJECT.
\end{itemize}
\textbf{Who is participating in the evaluation? }
\begin{itemize}
    \item \textbf{Automated:} N/A
    \item \textbf{Human:} Recruited 12 professional journalists (\emph{domain experts} \& \emph{intended users}) who worked in any medium (e.g., digital publications, newspapers, radio, TV) and were English speakers based in the US. Participants must have written press releases in the past. 
\end{itemize}
\textbf{When is evaluation conducted? }
\begin{itemize}
    \item \textbf{Time Scale}: Evaluation occured on the \emph{short-term} time scale. They took up to 60 minutes to complete and participants received \$30 for their time. Participants were shown a video demonstration of how to use the features in each system and then given 15 minutes with each tool to brainstorm a story idea
\end{itemize}
\textbf{How is evaluation validated?}
\begin{itemize}
    \item \textbf{Validation}: The tool and press release order were counterbalanced to prevent a learning effect (\emph{validity}). There's no mention of \emph{reliability} measures. 
\end{itemize}
\end{tcolorbox}
    \caption{Example \evalname evaluation card for AngleKindling~\cite{petridis2023anglekindling}.}
    \label{fig:angle-card}
\end{figure*}

\end{document}